\documentstyle[aps,prl,twocolumn,psfig]{revtex}
\begin{document}
\flushbottom
\draft
\twocolumn[\hsize\textwidth\columnwidth\hsize\csname
@twocolumnfalse\endcsname
\title{
Instability of antiferromagnetic magnons in strong fields
}
\author{M. E. Zhitomirsky$^{1,2}$ and A. L. Chernyshev$^{3,}$\cite{perm}}
\address{
$^1$Department of Physics, University of Toronto, Toronto M5S 1A7, Canada \\
$^2$L. D. Landau Institute for Theoretical Physics, Moscow 117334, Russia \\
$^3$Department of Physics, Queen's University, Kingston K7L 3N6, Canada
}
\date{\today}
\maketitle

\widetext\leftskip=1.9cm\rightskip=1.9cm\nointerlineskip\small
\begin{abstract}
\hspace*{2mm}
We predict that spin-waves in an ordered quantum antiferromagnet (AFM)
in a strong magnetic field
become unstable with respect to spontaneous
two-magnon decays. At zero temperature,
the instability occurs between the threshold field
$H^*$ and the saturation field $H_c$. As an example, we
investigate the high-field
dynamics of a Heisenberg antiferromagnet on a square lattice
and show that the single-magnon branch of the spectrum disappears
in the most part of the Brillouin zone.
\end{abstract}
]

\narrowtext

There are several reasons of studying the effect of strong magnetic
field on quantum AFMs. A growing family
of weakly interacting spin systems, which includes 
chain \cite{incom}, ladder \cite{ladder}, and square-lattice
AFMs \cite{Landee}, 
allows now experiments in previously unreachable
field regimes.
The high-field physics is proven to be rich for one-dimensional (1D)
AFMs, to mention only incommensurate gapless modes in a 
spin-$\frac{1}{2}$ chain \cite{incom}.
At the same time, in 2D or 3D the field
evolution of the N\'eel ordered ground state is trivial.
Spins cant gradually from the antiparallel structure 
until the magnetization saturates at the critical field 
$H_c$. Zero-point fluctuations vanish at the same field making
the ground state at $H>H_c$ purely classical.
One can suggest that similar quasi-classical scenario applies
also for the spin dynamics.
In contrast, we predict in this Letter that on the way to the saturated
phase the excitation spectrum of an ordered AFM undergoes
unexpectedly strong transformations.
In high fields below $H_c$, magnons are overdamped and disappear
in the most part
of the Brillouin zone (BZ).

The effect originates from the field-induced hybridization of 
single-magnon states with two-magnon continuum.
Previously, Osano {\it et al.\/}\ \cite{OSh} investigated 
the effect of such an interaction
on the two-magnon dynamical response at low fields $H\ll H_c$.
In this limit the hybridization leads to a weak renormalization
of the one-magnon spectrum.
In a strong field, however, two-magnon continuum overlaps
with the single-magnon branch and produces instability of the latter.
At $T=0$, there is a threshold field 
$H^*$ ($H^*\approx 0.76 H_c$ for square and cubic
lattices) above which magnons become unstable with respect to spontaneous
decays.
The argument for that is based on softening of the spin-wave velocity
of the sound mode near the AFM wave-vector $\bf Q$.
For small $\tilde{\bf k}=\bf k-Q$,
the magnon energy is approximately  
\begin{equation}
\omega_{\bf k}\approx c \tilde{k}(1+\alpha\tilde{k}^2) .
\label{lin}
\end{equation}
In zero field both the classical and renormalized dispersions 
bend downward implying $\alpha<0$ \cite{Oguchi,AFM-T,Canali}. 
On the other hand, at $H=H_c$ 
the low-energy asymptote for 
$\omega_{\bf k}$ coincides with the corresponding expression
for a ferromagnet: $\omega_{\bf k}\propto\tilde{k}^2$.
Simple continuity arguments predict that the high-field upward curvature
of the spectrum for $\tilde{k}\rightarrow 0$ 
changes to the low-field downward bend at some 
field $H^*<H_c$, where the cubic factor $\alpha$ vanishes.
Therefore, $\alpha$ is positive in the field region $H^*<H<H_c$
and long-wave magnons are kinematically unstable
towards the spontaneous decay \cite{Pit}.
 
The above argument for the existence of the threshold field $H^*$
is general and valid for all {\it ordered} quantum AFMs irrespectively
of their dimensionality, length of spin,  
or position
of $\bf Q$. In the following, we study in detail the high-field
dynamics of 
a spin-$\frac{1}{2}$ Heisenberg antiferromagnet with
nearest-neighbor interaction on a square lattice described by
the Hamiltonian
\begin{equation}
\hat{\cal H} = \sum_{\langle i,j\rangle} {\bf S}_i \cdot {\bf S}_j 
- H \sum_i S_i^{z_0} . 
\label{H}
\end{equation}
The ground state of this model is ordered at $0\le H\le H_c$
for arbitrary value of the on-site spin $S$ \cite{review,ZhNik}. 

For the canted AFM phase
\begin{equation}
\begin{array}{rcl}
S_i^{x_0} & = & S_i^ze^{i{\bf Q\cdot r}_i}\cos\theta+S_i^x\sin\theta ,
\\
S_i^{z_0}&=&S_i^z\sin\theta-S_i^x e^{i{\bf Q\cdot r}_i}\cos\theta  ,
\end{array}
\label{transf}
\end{equation}
$S_i^{y_0}=S_i^y$, and ${\bf Q}=(\pi,\pi)$ 
the boson representation of $\hat{\cal H}$ 
is obtained by applying the Dyson-Maleev 
transformation in the twisted frame:
$S_i^z=S-a^\dagger_ia_i^{_{}}$,  
$S_i^+=\sqrt{2S}(1-a^\dagger_ia_i^{_{}}/2S)a_i$, 
$S_i^-=\sqrt{2S}a^\dagger_i$.
The tilting angle
$\theta$ is determined from vanishing of the linear boson term
$\hat{\cal H}^{(1)}$: $\sin\theta = H/8S$, 
$H_c=8S$. Apart from a constant, the boson Hamiltonian is 
a sum of quadratic, cubic and quartic terms:
$\hat{\cal H} =  
\hat{\cal H}^{(2)} + \hat{\cal H}^{(3)} + \hat{\cal H}^{(4)}$ \cite{HP}.
Neglected terms containing products of five and six boson operators
are of higher orders in an expansion parameter $\cos^2\!\theta/2zS$
\cite{AFM-T},
$z$ being the number of nearest-neighbors,
and give small corrections even for $S=\frac{1}{2}$.

The quadratic Hamiltonian  
\begin{equation}
\hat{\cal H}^{(2)} = \sum_{\bf k}
\bigl[A_{\bf k}a^\dagger_{\bf k}a_{\bf k}^{_{}} 
-\case{1}{2}B _{\bf k} (a_{\bf k}a_{-\bf k}+a^\dagger_{\bf k}
a^\dagger_{-\bf k})\bigr]  ,
\label{H2}
\end{equation}
where $A_{\bf k}= 4S(1+\sin^2\!\theta\,\gamma_{\bf k})$,  
$B_{\bf k} = 4S\cos^2\!\theta\,\gamma_{\bf k}$, and 
$\gamma_{\bf k}=\frac{1}{2}(\cos k_x+\cos k_y)$, 
is diagonalized by the Bogoliubov transformation 
$a_{\bf k}=u_{\bf k}b_{\bf k}+v_{\bf k}^{_{}} b_{-\bf k}^\dagger$
with $u_{\bf k}^2+v_{\bf k}^2=A_{\bf k}/\omega_{\bf k}$,
$2u_{\bf k}v_{\bf k} = B_{\bf k}/\omega_{\bf k}$, giving  
the classical spin-wave energies 
\begin{equation}
\omega_{\bf k}=4S\sqrt{(1+\gamma_{\bf k})(1-\cos2\theta\,\gamma_{\bf k})} .
\label{cl}
\end{equation}
The magnon spectrum is defined in the paramagnetic BZ. 
It has a sound mode near $\bf k=Q$
with the classical spin-wave velocity $c=2\sqrt{2}S\cos\theta$.

In the leading order in $1/zS$ quantum corrections to the magnon
dispersion can be found from the Dyson equation for the normal
Green's function: 
$G({\bf k},t)=-i\langle T b^{_{}}_{\bf k}(t)b^\dagger_{\bf k}\rangle$, 
neglecting
anomalous contributions to the self-energy \cite{AFM-T}.
To find $\Sigma({\bf k},\omega)$ we express the
interaction terms 
via quasiparticle operators $b_{\bf k}$ and consider the first-order
perturbation from the quartic part 
and the second-order
perturbation from the cubic part. 

\begin{figure}[hp]
\vspace*{0.5cm}
\centerline{\psfig{figure=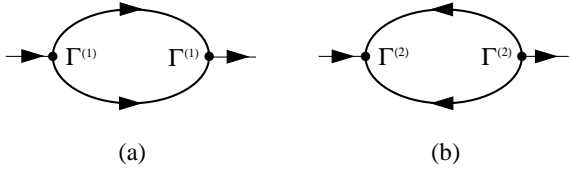,height=2.1cm,angle=0}}
\vspace{0.5cm}
\caption{Lowest-order diagrams contributing to the magnon self-energy.}
\end{figure}

There are two interaction terms with cubic vertices 
from $\hat{\cal H}^{(3)}$ \cite{ZhNik} 
\begin{eqnarray}
\hat{V}_1 &=& \frac{1}{2!}\sum_{1-2-3={\bf Q}} \Gamma^{(1)}_{1;23}
\bigl(b^\dagger_3b^\dagger_2b_1^{_{}} + {\rm h.\,c.}\bigr) , 
\nonumber \\
\hat{V}_2 &=& \frac{1}{3!}\sum_{1+2+3={\bf Q}} \Gamma^{(2)}_{123}
\bigl(b^\dagger_3b^\dagger_2b^\dagger_1 + {\rm h.\,c.}\bigr) ,
\end{eqnarray}
where we abbreviated momenta with their subscripts. 
The first term $\hat{V}_1$ is responsible for the 
hybridization between single-
and two-magnon spectra
and is the principal interaction of the problem.
The vertices are given by
$\Gamma^{(1)}_{1;23}=-\sqrt{8S}\sin\!2\theta\Phi({\bf k};{\bf q})$,
\begin{eqnarray}
\Phi(1;23) & = & \gamma_1(u_1+v_1)(u_2v_3+v_2u_3) 
+\gamma_2(u_2+v_2)(u_1u_3  \nonumber \\
& & \mbox{}+v_1v_3)+\gamma_3(u_3+v_3)(u_1u_2+v_1v_2) , 
\label{Phi}
\end{eqnarray}
and $\Gamma^{(2)}_{123}= -\sqrt{8S}\sin\!2\theta F({\bf k},{\bf q})$,
where similar expression for $F({\bf k},{\bf q})$ can 
be found in \cite{ZhNik}.
The lowest-order 
contributions to the self-energy shown in Fig.~1 are 
\begin{mathletters}
\begin{eqnarray}
\Sigma^{(1)}({\bf k},\omega) & = & 4S\sin^2\!2\theta\sum_{\bf q}
\frac{\Phi({\bf k};{\bf q})^2}{\omega-\omega_{\bf q}-\omega_{\bf k-q+Q}+i0} ,
\label{S1} \\
\Sigma^{(2)}({\bf k},\omega) & = & -4S\sin^2\!2\theta\sum_{\bf q}
\frac{F({\bf k},{\bf q})^2}
{\omega\!+\!\omega_{\bf q}\!+\!\omega_{\bf Q-k-q}\!-\!i0} .
\end{eqnarray}
\end{mathletters}
The first term describes a virtual decay of a magnon into 
two-particle intermediate states.
Frequency-independent contributions to the dispersion
arise from $\hat{\cal H}^{(4)}$
and from the renormalization of the tilting angle
\cite{ZhNik}:
\begin{eqnarray*}
&&\Sigma^{(3)}({\bf k})= 
4(u^2_{\bf k}+v^2_{\bf k})\bigl[
- n\cos 2\theta + \Delta\cos^2\!\theta - m\sin^2\!\theta \\
&& +\gamma_{\bf k}(-m\cos 2\theta
+\mbox{$\frac{1}{2}$}\delta\cos^2\!\theta-n\sin^2\!\theta)\bigr] 
-8u_{\bf k}v_{\bf k} 
\\
&& 
\times\!\bigl[\case{1}{2}\Delta\!\sin^2\!\theta\! 
-\!\case{1}{2}m\!\cos^2\!\theta\! 
+\!\gamma_{\bf k}(\Delta\!\cos 2\theta\! -\!
n\!\cos^2\!\theta\!+\!\mbox{$\frac{1}{2}$}\delta\!\sin^2\!\theta)\!\bigr],
\\
&&\Sigma^{(4)}({\bf k})=8\sin^2\!\theta(\Delta\!-\!n\!+\!m)
\bigl[\!
(u^2_{\bf k}\!+\!v^2_{\bf k})(1\!-\! \gamma_{\bf k}) \!
-\!2 u_{\bf k}v_{\bf k}\gamma_{\bf k}\!\bigr] , 
\end{eqnarray*}
where various two-boson contractions are defined as
$n=\sum_{\bf k}v_{\bf k}^2$,  
$m=\sum_{\bf k}v_{\bf k}^2\gamma_{\bf k}^{_{}}$,
$\delta=\sum_{\bf k}u_{\bf k}v_{\bf k}$,  
$\Delta=\sum_{\bf k}u_{\bf k}v_{\bf k} 
\gamma_{\bf k}$. 
Four-magnon vertices do not contribute to the dispersion to this
order in $1/zS$.

Perturbation theory gives the magnon energy with the first
$1/S$ corrections as
\begin{equation}
\label{w_ren}
\omega^{\rm pert}_{\bf k} = \omega_{\bf k} +
\sum_i \Sigma^{(i)}({\bf k},\omega_{\bf k}).
\end{equation}
Fig.~2 shows classical and renormalized spectra for $S=\frac{1}{2}$
in two representative
fields.
At $H=0$, the only non-vanishing correction
$\Sigma^{(3)}({\bf k})$ yields a 16\% enhancement
of magnon energies in the entire Brillouin zone \cite{Oguchi,Canali}.
In low fields the magnon spectrum resembles
zero-field result except the vicinity of the BZ center.
The mode at ${\bf k}=0$
describes a uniform precession of spins about $\bf H$ and,
for axially symmetric systems its
classical frequency $\omega_0=H$ is exact, i.e., not changed
by quantum effects \cite{Golos}.
With increasing field 
the cubic terms grow and push down magnon energies.
Above $0.7H_c$,
correction from $\Sigma^{(1)}({\bf k},\omega_{\bf k})$ exceeds 50\%
of $\omega_{\bf k}$
signifying a break down of the $1/S$ perturbation expansion for spin-wave frequencies 
(\ref{w_ren}). The 
correct renormalized spectrum plotted
in Fig.~2 is recovered by solving the Dyson equation
\begin{eqnarray}
\label{Dyson}
\omega-\omega_{\bf k} -\Sigma({\bf k},\omega)=0.
\end{eqnarray}
Standard $1/S$ expansion  breaks down for any finite $S$ when 
the bottom of two-magnon continuum
becomes nearly degenerate with a part of the single-magnon spectrum 
for $H$ close to the decay threshold field $H^*$.
Unrenormalized ``classical'' value of $H^*$ is calculated by expanding
Eq.~(\ref{cl}) up to the third order in $\tilde{k}$. The cubic factor
$\alpha$ is anisotropic and vanishes first for $\tilde{k}_x=\tilde{k}_y$
at $H^*=\frac{2}{\sqrt{7}} H_c\approx 0.756 H_c$.

At $H>H^*$ 
the magnon self-energy acquires imaginary part due to spontaneous
two-magnon decays, which obey the energy conservation law 
\begin{equation}
\omega_{\bf k} = \omega_{\bf q} + \omega_{\bf k-q+Q} .
\label{conserv}
\end{equation}
Spin-wave is stable, if Eq.~(\ref{conserv}) has only trivial solutions: 
$\bf q= k,Q$. In this case, the two-magnon
density of states $\rho_2({\bf k},\omega)= 
\sum_{\bf q} \delta(\omega-\omega_{\bf q}-\omega_{\bf k-q+Q})$ vanishes
at $\omega<\omega_{\bf k}$. 
$\rho_2({\bf k},\omega)$ 
has Van Hove singularities determined by symmetry extrema
$\bf k-q^*+Q=q^*+G$, where $\bf G$ is a reciprocal lattice vector.
At $H>H^*$, one of these points 
${\bf q}^*=({\bf k+Q})/2$ crosses the one-magnon state if
\begin{equation}
\omega_{\bf k} \ge 2\omega_{({\bf k+Q})/2}. 
\label{boundary}
\end{equation}
When applied to the phonon mode, Eq.~(\ref{boundary}) is equivalent
to the previously used condition $\alpha>0$
[note the difference between $k$ and $\tilde{k}$ in Eqs.~(\ref{lin})
and (\ref{boundary})]. At $H>H^*$, 
classical magnons (\ref{cl}) decay in the region around $\bf k=Q$
and remain stable in the vicinity of uniformly precessing mode ${\bf k}=0$
with the decay threshold boundary given by equality sign in 
Eq.~(\ref{boundary}). 

\begin{figure}[hp]
\vspace*{70mm}
\includegraphics{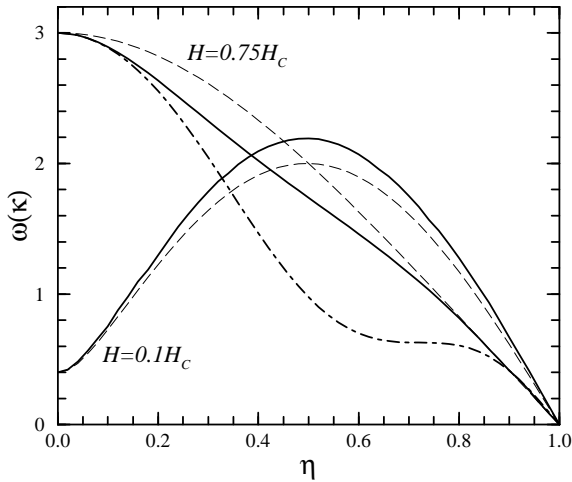}
\vspace*{-1mm}
\caption{Magnon dispersion for ${\bf k}=\pi(\eta,\eta)$: 
(i) thin dashed lines represent the classical spectrum $\omega_{\bf k}$; 
(ii) dot-dashed lines are $\omega^{\rm pert}_{\bf k}$; 
(iii) solid lines are solution of 
the Dyson equation. For $H=0.1H_c$, curves (ii) and (iii) 
are indistinguishable. 
}
\end{figure}
\vspace*{-2mm}

Pitaevskii argued that such boundaries are actually termination points
of the spectrum \cite{Pit}. This fact is a consequence of 
logarithmic divergence of the
one-loop diagram near the threshold:
$\Sigma^{(1)}({\bf k},\omega)\propto \ln(-\Delta\tilde{\omega}/R)$,
where $\tilde{\omega}=\Delta\omega+v_c\Delta k$,
$v_c$ is a velocity 
of created magnons, and $R$ is a cut-off. 
Termination point for the decay threshold into a pair of rotons is known
in the spectrum of superfluid $^4$He.
Since the self-energy is singular,
a proper renormalization of the cubic vertex
is necessary in that case \cite{Pit}.
Our problem, however, is
different, because two new magnons created in an
elementary decay process are again unstable.
Near the decay threshold, the dominant contribution to 
$\Sigma^{(1)}$ arises from the close vicinity of ${\bf q=(k+Q)}/2$,
therefore we can approximate 
${\rm Im}\,\omega_{\bf q}\approx{\rm Im}\,\omega_{\bf k-q+Q}=
\frac{1}{2}\Gamma$. Even small $\Gamma$'s remove completely
divergence of the diagram (Fig.~1a):
\begin{equation}
\Sigma^{(1)}({\bf k},\omega)\!\propto\! \Bigl[
\ln\frac{\sqrt{\Delta\tilde{\omega}^2\!+\!\Gamma^2}}{R}
-i\Bigl(\frac{\pi}{2}\!+\!
\arctan\frac{\Delta\tilde{\omega}}{\Gamma}\Bigr)\Bigr].
\label{Sigma}
\end{equation}
This analysis leads us to two conclusions:
(i) vertex corrections can be neglected, but
instead a self-consistent 
treatment of intermediate magnons and their decaying rates
is required; (ii) the boundary (\ref{boundary}) between 
stable and decay regions in BZ is smeared and at $H>H^*$ all magnons
simultaneously acquire finite life-times.

In the limit $\tilde{k}\rightarrow 0$ the decaying rate is small
and can be found analytically.
The self-consistent Born approximation (SCBA)
with dressed lines in Fig.~1a gives
\begin{eqnarray}
&&{\rm Im}\,\Sigma^{(1)}({\bf k},\omega)=-4\pi S\sin^2\!2\theta
\nonumber \\
&&\hspace*{4mm}\mbox{}\times \int \frac{d^2q}{(2\pi)^2} \Phi^2({\bf k};{\bf q})
\phi(\omega-\omega_{\bf q}-\omega_{\bf k-q+Q}) ,
\end{eqnarray}
where function $\phi(\omega)$ is normalized by
$\int\phi(\omega)d\omega=1$ and
has a characteristic width of the order of a sum of decay rates
of created magnons. 
Using the asymptotic form $\Phi({\bf k};{\bf q})\approx
-\frac{3}{4}[\tilde{k}\tilde{q}(\tilde{k}-\tilde{q})/
\sqrt{2}\cos^3\!\theta]^{1/2}$ we find
\begin{equation}
\omega_{\bf k}\approx c \tilde{k}(1+\alpha\tilde{k}^2
-i\beta\tilde{k}^2) ,
\end{equation}
with
$\beta = (9\sin^2\theta/64\pi\sqrt{2}\cos^3\theta)^{2/3}$.
Similar to a 3D case \cite{Pit}, this expression
for $\beta$ is valid beyond a narrow field region 
near $H^*$ determined by higher-order nonlinearities.

Another region in BZ, where
damping is small and excitations are well
defined, lies near the zone center. This follows from
the same arguments \cite{Golos},
which predict no effect on $\omega_0$
from quantum fluctuations,
since the exact frequency at ${\bf k}=0$ is
real and has no imaginary part. 

The spin-wave damping in the entire BZ is considered by
using the magnon spectral function $A({\bf k},\omega)
=-\frac{1}{\pi}{\rm Im}G({\bf k},\omega)$, where $G({\bf k},\omega)$
is calculated in the Born approximation.
Results for $A({\bf k},\omega)$ at $H=0.85H_c$ and
$H=0.9H_c$ are presented in Fig.~3(a) and 3(b), respectively.
In the non-self-consistent approximation (dashed lines), 
$A({\bf k},\omega)$
consists of a narrow one-magnon peak and two-magnon side-band,
which exhibits loose maximum at higher energies.
The quasiparticle peaks survive
even in the classical decay region
Eq.~(\ref{boundary}), because hybridization 
pushes them out from the two-magnon continuum.
This spurious feature arises due to the lack of self-consistency, as
the two sides of Eq.~(\ref{conserv}) are treated with 
different accuracy in the non-SCBA. 
That is, the magnon energy is renormalized while the energy of the continuum 
(right-hand side) 
is given by a classical expression.
Peaks disappear in the SCBA, which 
takes into account modification of two-particle
continuum due to renormalization of one-magnon states.
We performed numerical solution
of the Dyson equation in the SCBA with $96\times 96$ $\bf k$-points in BZ.
Results are plotted in Fig.~3 by solid lines.
Our analysis shows that a self-supporting instability 
intensifies at $H=0.85H_c$ 
for magnons from the region $k<Q/2$, which 
are classically stable
but become strongly damped since they lie in the
decay region for renormalized spectrum.

Magnons peaks in $A({\bf k},\omega)$ are strongly suppressed ($H=0.85H_c$) and
disappear ($H=0.9H_c$) 
because of the shift of single-particle pole in $G({\bf k},\omega)$
from the real axis into 
the complex plane  due to rapidly growing decay surface (\ref{conserv}).
This is in contrast with Pitaevskii's scenario
for superfluid $^4$He, where the pole's residue becomes vanishingly small \cite{Pit}.
Single-magnon excitations reappear again only at $H\approx 0.99H_c$, where the
decaying vertex $\Gamma^{(1)}(1;23)$ is small.

\begin{figure}[hp]
\vspace*{80mm}
\includegraphics{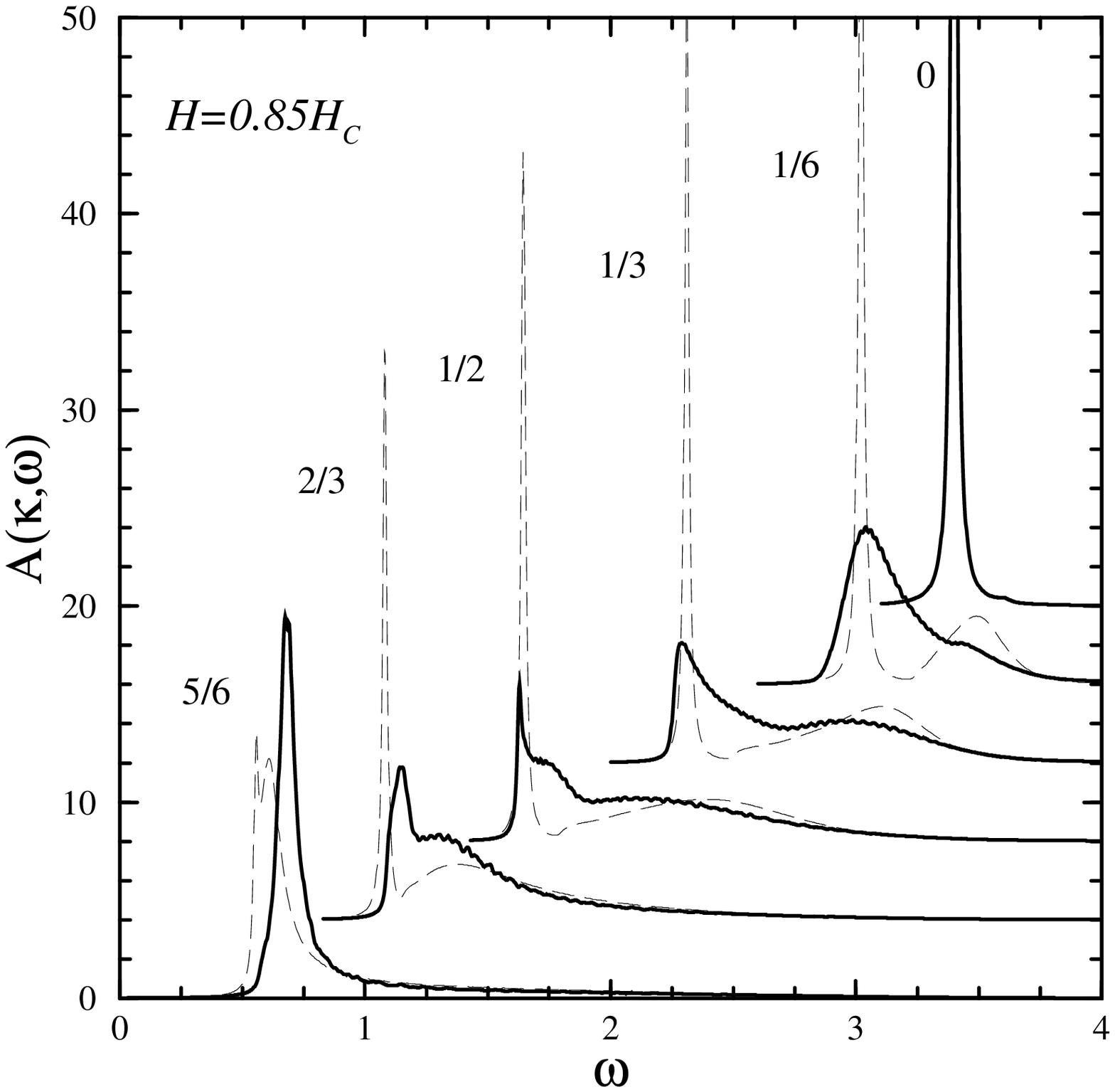}
\vspace*{74mm}
\includegraphics{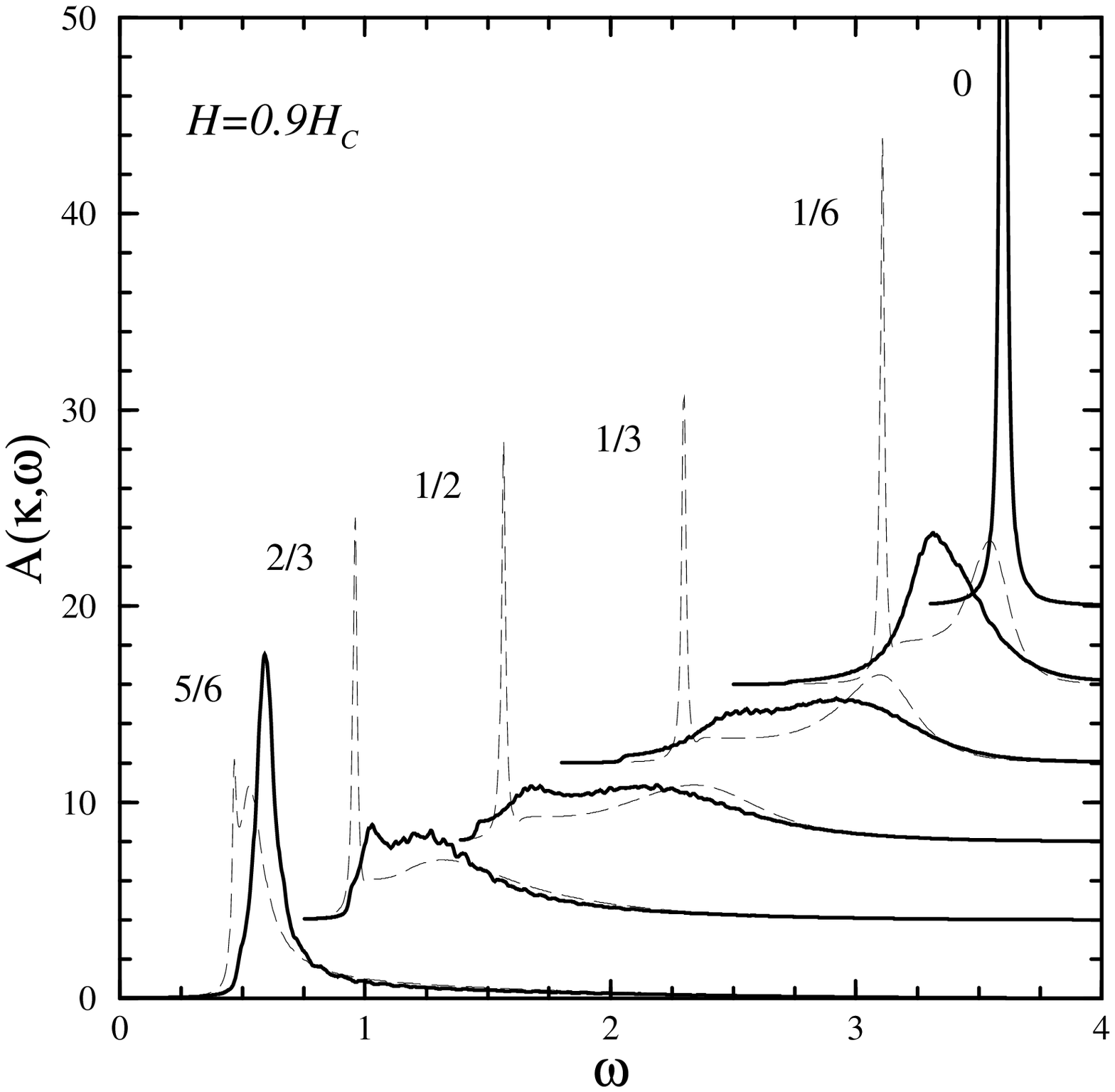}
\caption{Magnon spectral function for several points ${\bf k}=\pi(\eta,\eta)$,
(a) $H=0.85H_c$, (b) $H=0.9H_c$.
Thin dashed lines represent the
non-SCBA. Solid lines are the results of the
SCBA. Corresponding values of $\eta$ are shown near each curve.}
\end{figure}
\vspace*{-2mm}

To complete our study, we investigate now the high-field dynamical 
response of the square-lattice AFM, which is probed in
neutron diffraction experiments.
Our calculation of the dynamical structure factor
$S^{\alpha\beta}({\bf k},\omega)= \int dt \langle S^\alpha_{\bf k}(t)
S^\beta_{-\bf k}\rangle e^{i\omega t}$ is standard. We express
it at $T=0$ via the spin Green's function
$G^{\alpha\beta}({\bf k},t)= -i\langle TS^\alpha_{\bf k}(t)
S^\beta_{-\bf k}\rangle$  by 
$S^{\alpha\beta}({\bf k},\omega)= -2\theta(\omega)
G^{\alpha\beta}({\bf k},\omega)$.
The latter is related to the boson Green's
function by using Eq.~(\ref{transf}) and the Dyson-Maleev 
transformation.
The result for the inelastic part of the dynamical structure factor is
\begin{eqnarray}
&&S^{xx}({\bf k},\omega)=
\pi S\sin^2\!\theta(u_{\bf k}\!+\!v_{\bf k})^2\!A({\bf k},\omega) 
+\pi\cos^2\!\theta \nonumber \\
&& \hspace*{5mm}\times\sum_{\bf q}
(u_{\bf q}v_{\bf k-q+Q}\!+\!v_{\bf q}u_{\bf k-q+Q})^2 
\delta(\omega\!-\!\omega_{\bf q}\!-\!\omega_{\bf k-q+Q}), \nonumber \\
&&S^{yy}({\bf k},\omega)=\pi S(u_{\bf k}-v_{\bf k})^2 A({\bf k},\omega), \\
&&S^{zz}({\bf k},\omega)=
\pi S\cos^2\!\theta(u_{\bf k-Q}\!+\!v_{\bf k-Q})^2\!A({\bf k-Q},\omega) 
\nonumber \\
&& \hspace*{5mm} +\pi\sin^2\!\theta \sum_{\bf q}
(u_{\bf q}v_{\bf k-q}\!+\!v_{\bf q}u_{\bf k-q})^2 
\delta(\omega\!-\!\omega_{\bf q}\!-\!\omega_{\bf k-q}), \nonumber 
\end{eqnarray}
where we use the unperturbed spectrum for the two-magnon
contributions.
These two-magnon contributions described by the last terms in 
$S^{xx}$ and $S^{zz}$ are negligible in high fields because of 
small factors 
$v_{\bf k}\sim\cos^2\!\theta$. 
In the same approximation $u_{\bf k}\approx 1$ except for close vicinity
of $\bf Q$. Thus, in high fields $H^*<H<H_c$ the longitudinal component 
$S^{zz}({\bf k},\omega)$ reduces to its elastic part,
whereas transverse components 
$S^{xx}({\bf k},\omega)$ and $S^{yy}({\bf k},\omega)$
are proportional to $A({\bf k},\omega)$ with approximately same
momentum independent prefactors.

In conclusion, the nonlinear coupling between one-
and two-magnon states, which exists only 
in the canted AFM phase, becomes 
very important in high fields.
Together with the field-induced kinematic instability of the spectrum
this interaction
leads to suppression and disappearance of single-magnon excitations
for $H^*<H<H_c$.
Therefore, an intriguing
situation arises in the high field regime: the ground state
of an ordered quantum AFM is nearly classical, while
the excitation spectrum and the dynamical response are strikingly
different from the classical results. Though, our analysis
was restricted mainly to the spin-$\frac{1}{2}$ AFM in 2D, the qualitative
picture remains valid for other values of $S$ and for 3D.

We are grateful to D. I. Golosov, T. Nikuni, and V. I. Rupasov 
for useful discussions.
The work of A.L.C. was supported by the NSERC of Canada.

\vspace*{-5mm}

\end{document}